\documentclass[aps,prd,reprint,showpacs,superscriptaddress,floatfix]{revtex4-1}
\usepackage[T1]{fontenc} 
\usepackage{slashed}
\usepackage[utf8x]{inputenc}
\usepackage{color}
\usepackage[normalem]{ulem}
\usepackage{bm}
\usepackage{graphicx}

\usepackage{booktabs,amsmath}
\usepackage{amssymb}
\usepackage{array}
\usepackage{mathtools}
\usepackage{psfrag}

\usepackage{amsthm}
\usepackage{newlfont}
\usepackage{esdiff}
\usepackage{multirow}
\usepackage{graphics}
\usepackage{graphicx}
\usepackage{ae,aecompl,color}

\usepackage{bm}
\usepackage[bookmarks=true,bookmarksopen=true,bookmarksnumbered=true,bookmarksopenlevel=3]{hyperref}

\definecolor{airforceblue}{rgb}{0.36, 0.54, 0.66}
\definecolor{steelblue}{rgb}{0.27, 0.51, 0.71}
\definecolor{amber}{rgb}{1.0, 0.49, 0.0}
\begin{document}

\title{Exotic leptons at future linear colliders}
\author{\textsc{S.~Biondini}}
\affiliation{Physik-Department, Technische Universit\"{a}t M\"{u}nchen,
James-Franck-Str.~1, 85748 Garching, Germany}
\author{\textsc{O.~Panella}}
\affiliation{Istituto Nazionale di Fisica Nucleare, Sezione di Perugia, Via A.~Pascoli, I-06123 Perugia, Italy}
\email[({\bf Corresponding Author})\\ Email: ]{orlando.panella@pg.infn.it }

\date{\today}
\begin{abstract}
Doubly charged excited leptons give rise to interesting signatures for physics beyond the standard model at the present Large Hadron Collider. These exotic states are introduced in extended isospin multiplets which couple to the ordinary leptons and quarks either with gauge or contact effective interactions or a combination of both. In this paper we study the production and the corresponding signatures of doubly charged leptons at the forthcoming linear colliders and we focus on the electron-electron beam setting. In the framework of gauge interactions, the interference between the $t$ and $u$ channel is evaluated that has been neglected so far. A pure leptonic  final state is considered ($e^{-} \, e^{-} \rightarrow e^{-} \, e^{-} \, \nu_{e} \, \bar{\nu}_{e}$) that experimentally translates into a like-sign dilepton and missing transverse energy signature. We focus on the standard model irreducible background and we study the invariant like-sign dilepton mass distribution for both the signal and background processes. Finally, we provide the 3 and 5-sigma statistical significance exclusion curves in the model parameter space. We find that for a doubly charged lepton mass $m^*\approx 2 $ TeV the expected lower bound on the compositeness scale at CLIC, $\Lambda > 25$ TeV, is much stronger than the current lower bound from LHC  ($\Lambda > 5$ TeV) and remains highly competitive with the bounds expected from the run II of the LHC.
\end{abstract}

\maketitle

\section{Introduction}
\label{intro}
The Standard Model (SM) of particle physics is nowadays well tested within a plethora of low and high energy experiments. The recent discovery of a particle compatible with the missing fundamental scalar \cite{Chatrchyan:2012aa}, \cite{Aad:2012aa}, the Higgs boson, establishes the last great success of the SM. Of course the theoretical and experimental frontiers are just moved onto some further questions and issues. Among those, let us consider the so-called hierarchy problem or the question why the Higgs boson mass is light after including loop corrections. Indeed, the Higgs mass receives quantum corrections from all heavy SM particles (Higgs, gauge bosons and top quark). Such divergences are quadratic in a generic cutoff scale and not protected by any chiral symmetry like the fermion masses are. Moreover, any possible new physics scale laying between the electroweak and the Planck scale might push up the cutoff scale. Unless a fine tuning is considered as a satisfactory answer, a deeper understanding is desirable. 

Certainly several beyond the standard model scenarios  have been considered in the literature. Supersymmetry~\cite{Nilles:1984aa,Haber:1985aa}, extra-dimensions~\cite{Randall:1999ee,Appelquist:2001aa,Giudice:1999aa}, partial compositeness~\cite{Giudice:2007fh} and compositeness~\cite{Eichten:1983hw,Cabibbo:1983bk,Baur:1989kv} are just a few of them.  An active search for supersymmetric particles has been carried out at the Large Hadron Collider (LHC) with no positive outcome until now~\cite{Aad:2014iza},\cite{Aad:2014vma}. This brings to a quite narrow window for the minimal supersymmetric model even though many non-minimal realizations are still under investigation and not excluded by the current data. The compositeness approach is somewhat complementary to supersymmetry and no experimental evidence has been provided either. New unknown particles are supposed to be the constituents of the Higgs boson and its mass is induced by some internal dynamics as a meson mass is generated by an interacting quark anti-quark pair \cite{Kaplan:1983fs}. On the same footing, composite models also concern about the proliferation of the SM leptons and quarks collected in the three different generations. One of the main consequence of a further substructure is the possibility to observe excited fermions that would produce signatures to be studied at colliders \cite{Eichten:1983hw}. 
Indeed, many experimental analysis at colliders have been performed about the direct or indirect search of excited fermions, especially in the framework of four-fermion contact interactions. The stronger bounds are nowadays provided by the LHC experiments \cite{Nakamura:2010zzi},\cite{Khachatryan:2010jd}.

In this paper, we focus on excited leptons and in particular on those having an exotic charge $Q=2e$. These states appear in the context of phenomenological models that rely on isospin invariance and magnetic type interaction \cite{Cabibbo:1983bk}. Doubly charged leptons appear also in string inspired models \cite{Cvetic:2011iq} and in supersymmetric extensions of left-right symmetric models \cite{Demir:2008wt},\cite{Franceschini:2013aa}. Not prompt decaying doubly charged leptons have been also considered as a viable candidate for cold dark matter \cite{Fargion:2005ep}. 

The production cross section and relevant signatures for the excited doubly charged leptons have been studied in some detail for both the contact and gauge interactions in the LHC framework~\cite{Biondini:2012ny,Leonardi:2014aa}. In this work, we address the phenomenology of doubly charged leptons in the context of the forthcoming electron-positron linear colliders. In particular, we aim to investigate the signatures produced by the doubly charged leptons at the linear colliders facilities and the interplay with the corresponding SM  background. The International Linear Collider (ILC) \cite{Djouadi:2007ik} and the Compact Linear Collider (CLIC)~\cite{Linssen:2012hp,Accomando:2004sz} proposals  offer the possibility to have $e^{-} e^{-}$ beams. An interesting feature of the $e^{-} e^{-}$ scattering is its particular sensitivity to lepton violating processes, such as those that would be induced by the existence of heavy Majorana neutrinos \cite{London:1987nz}. Here we will  focus on this particular beam option, besides the standard one ($e^{+}e^{-}$), because it allows for the single production channel. Indeed, at tree level, the $e^{+}e^{-}$ configuration allows only pair production of the excited leptons and the available phase space would be very limited since the excited lepton mass is expected to be quite large. We consider a pure leptonic final state as follows: 
\begin{equation}e^{-} e^{-} \rightarrow E^{--} \nu_{e} \rightarrow W^{-}e^{-} \nu_{e} \rightarrow (e^{-}e^{-}) \bar{\nu}_{e}\nu_{e}.\end{equation}  
Since we want to shape the main features of the process we fix the flavour of the final state leptons within the first generation (electron flavour).

We emphasise that the ATLAS and CMS collaborations have recently provided  lower bounds on the mass of the excited leptons and of  long lived multi-charged particles.
Excited leptons have been searched in the process $p p \to \ell^*\ell \to \ell \ell \gamma$, assuming that the new particles are produced via contact interactions and decay via gauge interactions~\cite{Aad:2013jja,Chatrchyan:2013ac,Aad:2012ab}. They provide a lower bound on the mass of the excited leptons $m^* > 2.2 $ TeV for the special case in which the compositeness scale $\Lambda=m^*$.
When $\Lambda$ and $m^*$ are treated as independent parameters exclusion curves are given in the plane ($m^*,\Lambda$). For example for $m^*=0.6$ TeV the authors of ref.~\cite{Chatrchyan:2013ac} find $\Lambda > 10$ TeV.

As regards the long lived multi-charged particles  the ATLAS collaboration excludes particles up to 430 GeV based on the run at $\sqrt{s}=7$ TeV and with $L=5$ fb$^{-1}$ \cite{Aad:2013aa}. Similarly the CMS collaboration sets lower mass limits up to 685 GeV within the run at $\sqrt{s}=8$ TeV and with $L=18.8$ fb$^{-1}$ \cite{Chatrchyan:2013ab}. These exclusion limits do not directly apply to our model because Drell-Yan-like pair production is assumed for the multi-charged particle in the experimental analysis and the doubly charged lepton considered here promptly decays. The production of excited fermions at linear electron-positron and electron-proton colliders has been addressed already long ago \cite{Hagiwara:1985wt},\cite{Boudjema:1992em}. A more recent phenomenological study on excited fermions at linear colliders can be found in \cite{Cakir:2007wn}, where particles with spin $1/2$ and $3/2$ have been considered. 

In Sec.~\ref{sec_model} we briefly discuss the model where the doubly charged leptons are introduced. In Sec.~\ref{Sec2} we address the production cross section for the doubly charged lepton in the context of gauge mediated interactions, whereas Sec.~\ref{Sec3} is dedicated to the production cross section via contact interactions. In Sec.~\ref{Sec4} we study the SM irreducible background and we carry out the analysis of the relevant kinematic distributions for the signal and the SM background processes. Finally we summarize our conclusions and outlook in Sec.~\ref{Sec5}.  
\begin{figure}[t]
\centering
\includegraphics[width=8cm]{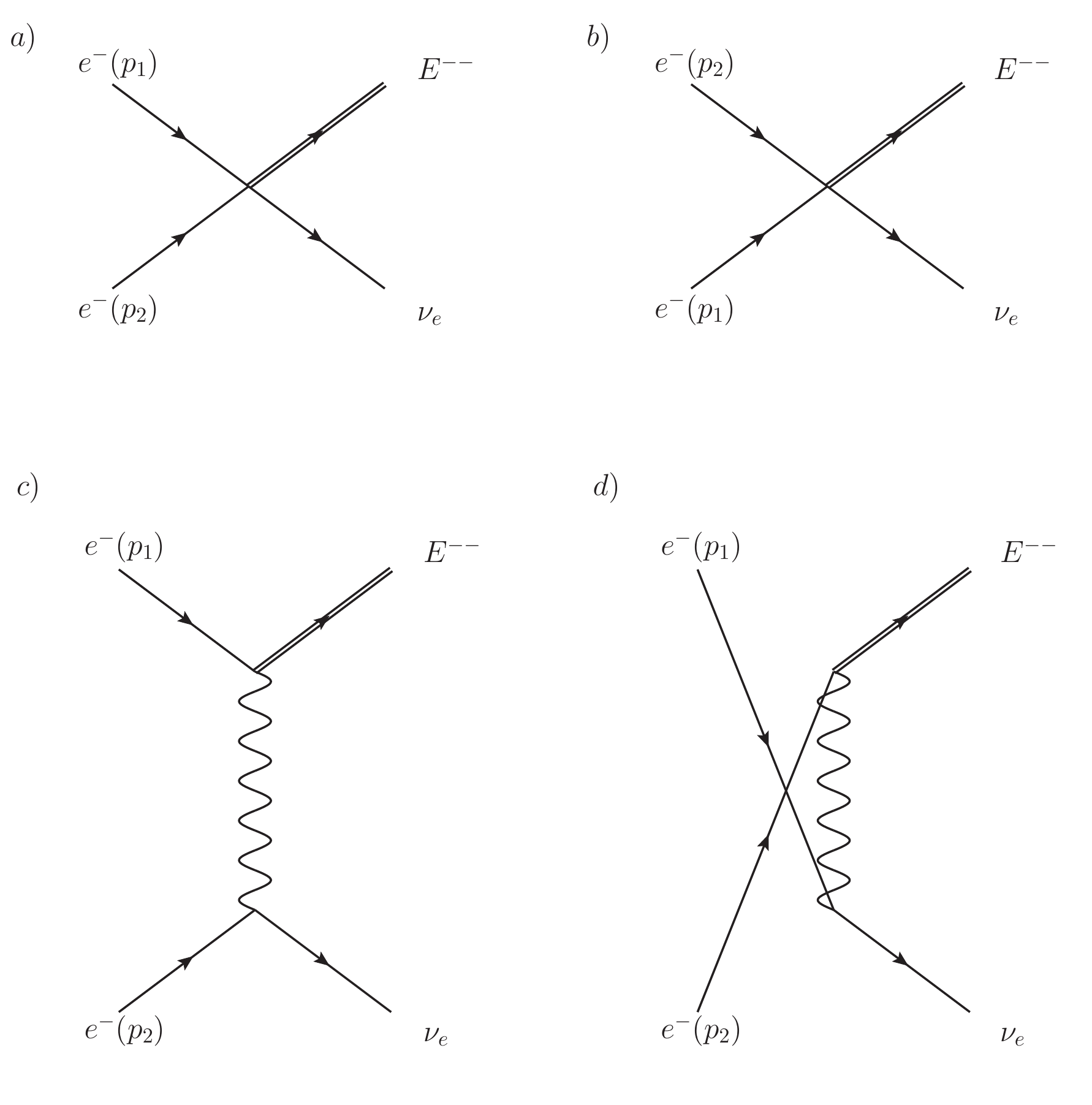}
\caption{\label{fig:diagrams} Diagrams for the the process $e^{-} \, e^{-} \rightarrow E^{--} \nu_{e}$ 
induced by the contact interaction Lagrangian ($a$ and $b$) and gauge interaction Lagrangian ($c$ and $d$). Solid lines stand for electrons and electron neutrinos, wiggled lines for the $W$ boson and solid double lines for the doubly charged lepton. We label with different momenta, $p_1$ and $p_2$, the incoming electrons.}
\end{figure}

\section{Extended Isospin Model}
\label{sec_model}

It is well known that in hadronic physics the strong isospin symmetry allowed to discover  baryon and meson resonances well before the observation of quarks and gluons. The properties of the hadronic states could be delineated using the SU(2) and SU(3) symmetries. In analogy with this it may be expected that, for the electroweak sector, the weak isospin spectroscopy could reveal some properties of excited fermions without reference to a particular internal structure.

The standard model fermions have $I_W=0$ and $I_W=1/2$ and the electroweak bosons have $I_W=0$ and $I_W=1$, so, combining them, we can consider fermionic excited states with $I_W\leq 3/2$. The multiplets with $I_W=1$ (triplet) and $I_W=3/2$ (quadruplet) include the doubly charged leptons that are studied in this work: 
\[ L_1 = \left( \begin{array}{l}
L^{0} \\
L^{-} \\
L^{--} \end{array} \right) ,
\qquad
L_{3/2} = \left( \begin{array}{l}
L^{+} \\
L^{0} \\
L^{-} \\
L^{--} \end{array} \right) \, ,\] 
with similar multiplets for the antiparticles.
While referring to the original work in \cite{Pancheri:1984sm} for a detailed discussion of all couplings and interactions, we discuss  here only the main features of the higher multiplets and write down   the relevant effective lagrangian density.  We refer to \cite{Biondini:2012ny} for further details and here we mention only  that the higher isospin multiplets ($I_W=1,3/2$) contribute solely to the iso-vector current and do not contribute to the hyper-charge current. Therefore the particles of these higher multiplets interact with the standard model fermions only via the $W$ gauge field. For the exotic doubly charged lepton of the  $I_{W}=1$ triplet and the one of the $I_{W}=3/2$ quadruplet the relevant interaction lagrangians  are respectively:
\begin{subequations}
\label{LagGI}
\begin{align}
\label{lagG1}
\mathcal{L}_{\text{G}}^{(1)}=i\frac{g\, f}{\Lambda}\left( \bar{\psi}_E \, \sigma_{\mu \nu} \, \partial^{\nu} \, W^{\mu} \, P_R \, \psi_e \right)  + h. c.\\
\label{lagG32}
\mathcal{L}_{\text{G}}^{(3/2)}=
 i\frac{g\,\tilde{f}}{\Lambda}\left( \bar{\psi}_E\sigma_{\mu \nu} \, \partial^{\nu} \, W^{\mu} \, P_L \, \psi_e  \right)  + h.c.
 \end{align}
\end{subequations}
where $f$, $\tilde{f}$ are couplings that parametrize the effective interaction for the $I_{W}=1$ and $I_{W}=3/2$, $g$ is the SU(2) coupling, $P_{L}=(1-\gamma^{5})/2$ and $P_{R}=(1+\gamma^{5})/2$ are the chiral projectors, $\sigma^{\mu \nu}=i\left[\gamma^{\mu},\gamma^{\nu} \right] /2$. The field $\psi_{E}$ stands for the doubly charged lepton either for the case $I_{W}=1$ or $I_{W}=3/2$, $\psi_{e}$ and $\psi_{\nu}$ are the electron and electron neutrino field respectively. The constants $f$ and $\tilde{f}$ are usually set to one in the literature and we keep this choice as well. The effective Lagrangian in (\ref{LagGI}) is made of dimension five operators and hence one inverse power of the new physics scale $\Lambda$ is there.
Let us now discuss contact interactions (CI). 

Contact interactions describe an effective vertex by a four-fermion interaction which is obtained after the high energy modes of order of the compositeness scale $\Lambda$  have been integrated out. The underlying theory is indeed not specified and an effective Lagrangian is set up in order to study the interaction between excited fermions and SM particles. We consider the general contact interaction Lagrangian to be 
\begin{equation}
\mathcal{L}_{CI}=\left( \frac{g^{2}_{*}}{2\Lambda^{2}}\right) j^{\mu} j_{\mu} \, , 
\end{equation}
where the current reads \cite{Baur:1989kv}
\begin{eqnarray}
j_{\mu}&=&\left(   \eta \bar{f}_{L} \gamma_{\mu} f_{L} +  \eta' \bar{f}_{L} \gamma_{\mu} f^{*}_{L} + \eta'' \bar{f^{*}}_{L} \gamma_{\mu} f^{*}_{L} + h.c. \right) \nonumber \\ &&+ (L \, \rightarrow \, R) \, ,
\label{effcontact}
\end{eqnarray} 
where $f_L$ stands for a SM fermion and $f^{*}_L$ for an excited fermion.
The constants in front of each vector current are usually put equal to one in the literature. In the following we do not consider the right-handed Lagrangian term in (\ref{effcontact}) for simplicity. The resulting effective Lagrangian in (\ref{effcontact}) is made of dimension six operators and hence two inverse powers of the new physics scale $\Lambda$ appear. According to the processes $a$ and $b$ displayed in Fig.~\ref{fig:diagrams}, we need the following current:
\begin{equation}
j_{\mu}= 
\left[ \bar{\psi}_{\nu}(x) \gamma_{\mu} P_{L} \psi_{e}(x) + \bar{\psi}_{E}(x)\gamma_{\mu} P_{L} \psi_{e}(x) + h.c. \right]   \, ,
\end{equation}   
where the fields $\psi_{\nu}$, $\psi_{e}$ and $\psi_{E}$ are the electron neutrino, electron and excited lepton respectively, whereas  $P_{L}=(1-\gamma^{5})/2$ is the chiral projector. The corresponding Lagrangian reads
\begin{equation}
\label{LagCI}
\mathcal{L}_{CI}= \frac{g^{2}_{*}}{\Lambda^{2}} \, \left[  \bar{\psi}_{\nu}(x) \gamma^{\mu} P_{L} \psi_{e}(x) \, \bar{\psi}_{E}(x)\gamma_{\mu} P_{L} \psi_{e}(x) + h.c.  \right] \, .
\end{equation}

%

The gauge interactions in Eqs.~(\ref{LagGI}) were implemented in {CalcHEP}~\cite{Pukhov:1999,Belyaev20131729} in \cite{Biondini:2012ny} with the help of {FeynRules}~\cite{Christensen:2008py}, a {Mathematica}~\cite{math} package which allows to write down the Feynman rules of any quantum field theory model described by a given Lagrangian.  In \cite{Leonardi:2014aa} the contact interactions in Eq.~\eqref{LagCI} have been implemented  in the same CalcHEP model of \cite{Biondini:2012ny}. Contact interactions have to be entered "by hand" in CalcHEP  with the help of an auxiliary gauge field~\cite{Belyaev20131729} which is exchanged by the fermion currents. Once this was accomplished the CalcHEP generator has been used   in \cite{Leonardi:2014aa}  to  address at LHC the interplay of gauge interactions, Eq.~\eqref{LagGI}, with contact interactions, Eq.~\eqref{LagCI}, in the phenomenology of the exotic states with respect to  \textit{single production cross sections} and the excited particles \textit{decays}.  Here we plan to address the same phenomenology aspects but with a focus on a Linear Collider facility.  
\begin{figure}[t]
\centering
\includegraphics[width=8cm]{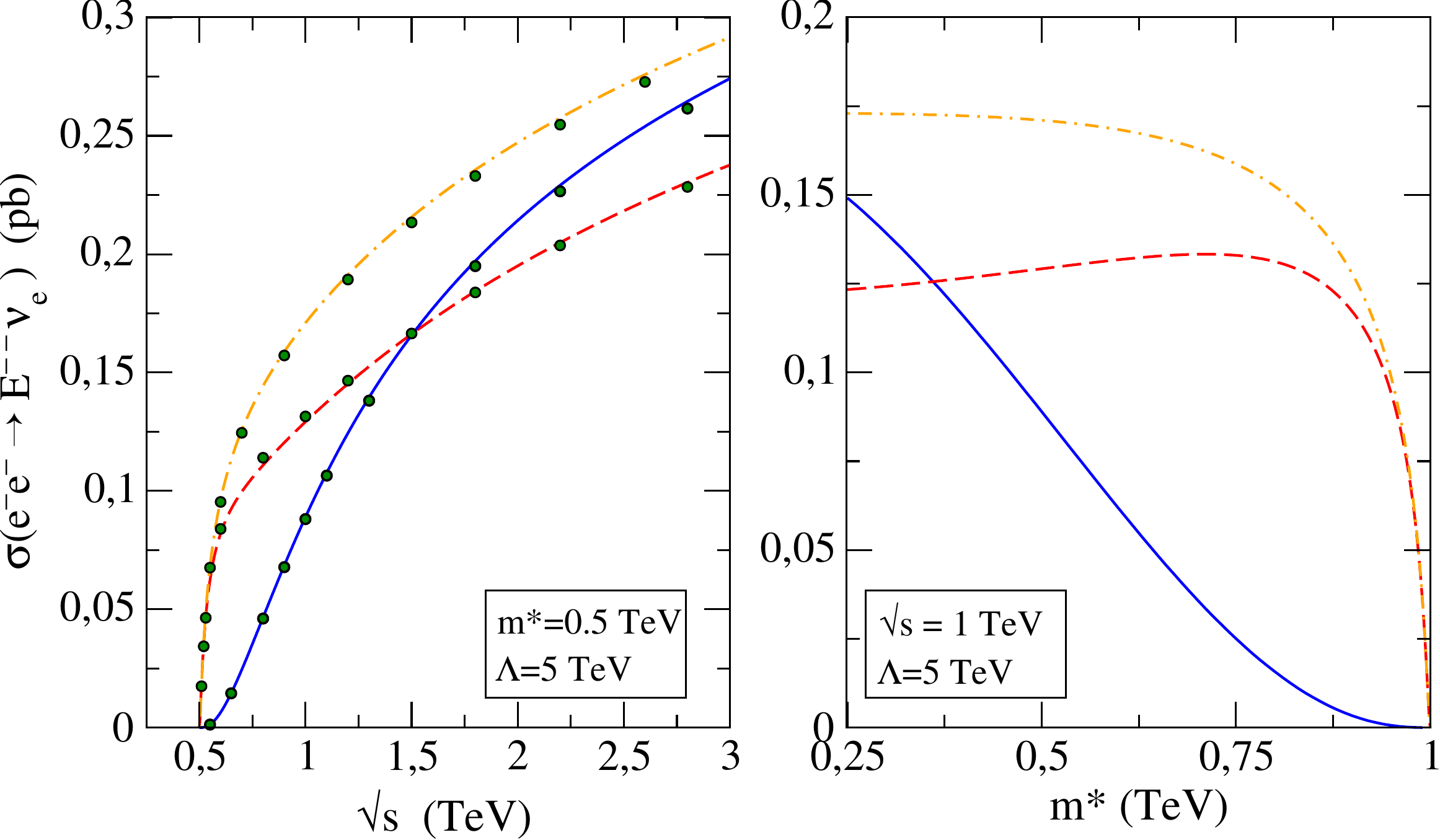}
\caption{\label{fig:interference} (Color online). The plot shows the total cross sections for the process $e^{-} e^{-} \to E^{--} \nu_{e}$. In the left panel the total cross section against the center of mass energy is displayed for the different isospin multiplets. The solid blue line stands for $E^{--} \in I_{W}=1$, the dashed red one for $E^{--} \in I_{W}=3/2$ with the interference contribution taken into account. At variance, the orange dot-dashed curve is the result when the interference is not considered. The right panel shows the total cross section against mass. Also here the curves for different multiplets are displayed. The compositeness parameters are set to ($m^{*}=0.5$ TeV, $\Lambda=5$ TeV). The green dots stand for the CalcHEP output.}
\end{figure} 

\section{Production cross sections} 
Excited leptons were first introduced in the context of compositeness and weak isospin invariance \cite{Cabibbo:1983bk}. Higher isospin multiplets, namely $I_{W}=1$ and $I_{W}=3/2$, were added to the standard ones ($I_{W}=0$ and $I_{W}=1/2$) in \cite{Pancheri:1984sm}. Exotic electromagnetic charges for the fermions are then allowed ($Q=4/3 \,e,5/3 \, e$ for quarks and $Q=2e$ for leptons). Such exotic charges, not present in the SM,  may lead to interesting signatures that can be investigated at colliders. In particular we will focus on the single production of the doubly charged excited electron $E^{--}$ at the $e^-e^-$ option of the linear collider, $e^{-} e^{-} \rightarrow E^{--} \nu_{e} $.

\subsection{Gauge interactions}
\label{Sec2}
In this subsection we focus on the production cross section for doubly charged leptons via gauge interactions. 
At variance with contact interactions, the gauge interactions enrich the phenomenology with the sensitivity to angular distributions and to different weak isospin multiplets. As a drawback, the production cross sections derived from gauge interactions are rather smaller than the ones predicted by contact interactions for the same value of the parameters $(m^{*},\Lambda)$. Here, $m^{*}$ is the mass of the excited doubly charged lepton and $\Lambda$ the compositeness scale.  

In the following, we show that the production cross section for the doubly charged lepton is different depending on $E^{--} \in I_{W}=1$ or $I_{W}=3/2$. The production cross sections of excited fermions via gauge interactions for the different multiplets were already evaluated in \cite{Pancheri:1984sm}. In particular the process $q_a \bar{q}_b \rightarrow f \, f^{*}$ has been addressed, where $q_a$ and $q_b$ are two generic SM quark and anti-quark, $f$ stands for a SM fermion and $f^{*}$ for an excited fermion. The authors give general expressions for the process at the parton level and they neglect the interference between the different kinematic channels. Here, we add the study of the interference between the $t$ and $u$ channel depicted in Fig. \ref{fig:diagrams}, diagrams $c$ and $d$ respectively, relevant for the present study. Indeed, the $t$-$u$ interference vanishes in the $I_{W}=1$ case but it plays a role when considering $I_{W}=3/2$. The difference is determined by the chiral projector involved in the vertex that couples the excited fermion to a SM one.

In the following, we express the differential cross sections in terms of the Maldestam variables. According to Fig.~\ref{fig:diagrams} one has to take into account the $t$ and $u$ channels and the interference amongst the two. We list the result as follows
\begin{widetext}
\begin{eqnarray}
\left( \frac{d \sigma}{d t}\right)_{I_{W}=1} &=& \frac{1}{4 s^2 \Lambda^2} \frac{g^4 f^2}{16 \pi} \frac{t}{\left( t-M^{2}_{W} \right)^2} \, \left[ m^{*2} (t-m^{*2}) +2su+m^{*2}(s-u)\right]  \nonumber
\\
&&+\frac{1}{4 s^2 \Lambda^2} \frac{g^4 f^2}{16 \pi} \frac{u}{\left( u-M^{2}_{W} \right)^2} \, \left[ m^{*2} (u-m^{*2}) + 2st + m^{*2}(s-t)\right] \, ,
\label{cross1}
\end{eqnarray}
\begin{eqnarray}
\left( \frac{d \sigma}{d t}\right)_{I_{W}=3/2} &=& \frac{1}{4 s^2 \Lambda^2} \frac{g^4 \tilde{f}^2}{16 \pi} \frac{t}{\left( t-M^{2}_{W} \right)^2} \, \left[ m^{*2} (t-m^{*2}) +2su-m^{*2}(s-u)\right]  \nonumber
\\
&&+\frac{1}{4 s^2 \Lambda^2} \frac{g^4 \tilde{f}^2}{16 \pi} \frac{u}{\left( u-M^{2}_{W} \right)^2} \, \left[ m^{*2} (u-m^{*2}) + 2st - m^{*2}(s-t)\right] \nonumber \, 
\\
&&+\frac{1}{8 s^2 \Lambda^2} \frac{g^4 \tilde{f}^2}{16 \pi} \frac{1}{\left( u-M^{2}_{W} \right)}\frac{1}{\left( t-M^{2}_{W} \right)} \left( 2stu + \frac{3}{4}utm^{*2}\right)  \, .
\label{cross3}
\end{eqnarray}
where $M_W$ is the $W$ boson mass. Let us remark some features of the above expressions. One may obtain the cross section for the $u$ channel from the $t$ one just by crossing symmetry, namely $t \leftrightarrow u$. Therefore, the second line in eqs. (\ref{cross1})-(\ref{cross3}) is obtained from the first one by performing the exchange between the Maldestam variables $t$ and $u$. The cross sections for the different isospin multiplets differ from the third term in the square brackets, flip of sign, and the interference term that is only present in eq. (\ref{cross3}). The differences are inherited from the chiral projector that enters the Lagrangian in (\ref{LagGI}), $P_{R}$ for $E^{--} \in I_{W}=1$ and $P_{L}$ for $E^{--} \in I_{W}=3/2$. We compute explicitly the interference term and we show that it accounts for a relevant effect in the integrated cross section for $E^{--} \in I_{W}=3/2$ as shown in Fig.~\ref{fig:interference}. Indeed, the cross section is reduced  by a factor of almost one third at $\sqrt{s}=1$ TeV and the effect becomes larger at higher center of mass energies. The result for the integrated total cross section reads as follows
\begin{equation}
\sigma_{I_{W}=1}=\frac{\alpha^2 f^2 \pi \left( s-m^{*2}\right) }{s^2 \Lambda^2 \sin ^4 \theta_w}  \left[ -2\left( s-m^{*2}\right) + \left( s-m^{*2} +2M^{2}_{W}\right) \ln \left( 1+ \frac{s-m^{*2}}{M^2_{W}}\right) \right]  \, , 
\label{cross1T}
\end{equation}  
\begin{eqnarray}
\sigma_{I_{W}=3/2}&=& \frac{\alpha^2 \tilde{f}^{2} \pi }{s \Lambda^2 \sin ^4 \theta_w}  \left[ -2s + m^{*2} \left( 1+\frac{M^2_W}{s-m^{*2}+M^{2}_W}\right)  +  \left( s+2M^{2}_{W}\right) \ln \left( 1+ \frac{s-m^{*2}}{M^2_{W}}\right) \right] \nonumber
\\
&&+ \frac{\alpha^2 \tilde{f}^{2} \pi }{8 s^2 \Lambda^2 \sin ^4 \theta_w}  \left( 8s+3m^{*2}\right) \left[  -s+m^{*2} +2M^{2}_{W} \ln \left( 1+ \frac{s-m^{*2}}{M^2_{W}}\right) \right]   \, ,
\label{cross3T}
\end{eqnarray}
\end{widetext}
where we introduce the sine of the Weinberg angle ($\sin \theta_w$) in order to have the QED coupling constant in the expressions.  
Their behaviour near threshold, $s \to m^{*2}$, may be parametrized at leading order in $x$ as follows:
\begin{equation}
\sigma_{I_{W}=1} \sim K \frac{x^4}{m^{*4}M^4_{W}}, \quad \sigma_{I_{W}=3/2} \sim \widetilde{K} \frac{x}{m^{*2}} ,
\end{equation}
where we define $x=s-m^{*2}$, $K=(\alpha^2 f^2 \pi)/(6\Lambda^2 \sin^4 \theta_W)$ and $\tilde{K}=(11\alpha^2 \tilde{f}^2 \pi)/(8\Lambda^2 \sin^4 \theta_W)$. The result for the total cross section for the different multiplets is crosschecked with the CalcHEP \cite{Belyaev20131729} output of the model in Fig.~\ref{fig:interference}.

\begin{table*}[t]
\begin{ruledtabular}
\begin{tabular}{ccccc}&&Base Kinematic Cuts&&\\\hline
$p_T^{min}(e^{-}) > 15 $ GeV&$p_T^{max}(e^{-}) > 50 $ GeV   &$p_T(\nu) > 15$ GeV&  $|\eta (e^{-})|<2.5$ & $\Delta R (e^{-},e^{-})>0.5$
\end{tabular}
\end{ruledtabular}
\caption{\label{Tab1} The base kinematic cuts adopted  in order to obtain the upper panel in Fig.~\ref{fig:invmass}. This choice is based on existing models of general purpose detectors for a linear collider~\cite{Djouadi:2007ik,Linssen:2012hp,Accomando:2004sz}. }
\end{table*}
\begin{table*}
\begin{ruledtabular}
\begin{tabular}{ccc}&Improved Kinematic Cuts&\\\hline
 $p_T^{max}(e^{-}) > 200 $ GeV   & $-1<\eta^{max} (e^{-})<2.5$ & $\slashed{E}_T>100$ GeV
\end{tabular}
\end{ruledtabular}
\caption{\label{Tab2} The improved kinematic cuts used in addition to the base cuts in order to obtain the lower panel in Fig.~\ref{fig:invmass}. Unchanged base cuts from Table~\ref{Tab1} are not repeated here. }
\end{table*}

\subsection{Contact interactions}
\label{Sec3}
In this subsection we show the results for the production cross section in the case of contact interactions.

\begin{figure}
\centering
\includegraphics[width=8cm]{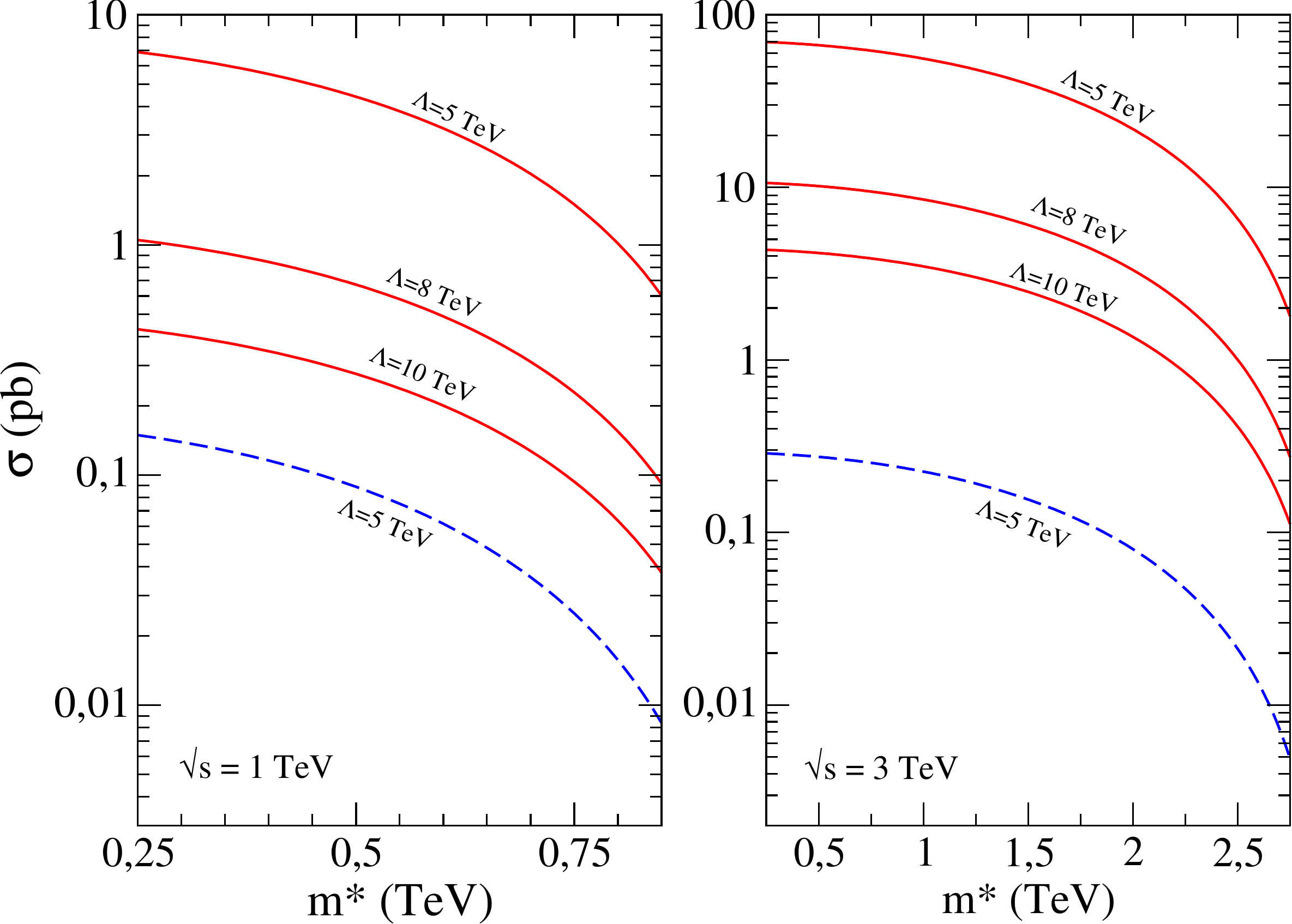}
\caption{\label{fig:crosssections} (Color online). The plot shows the total production cross section as a function of the excited lepton mass for different values of the compositeness scale $\Lambda$. The left panel refers to 1 TeV for the energy in the center of mass (highest expected energy for the ILC) whereas in the right panel we set the collision energy at 3 TeV (highest expected energy for CLIC). Contact interactions (solid lines) dominate over the gauge interactions (dashed line) for the production of the doubly charged lepton. We use $E^{--} \in I_{W}=1$ for the gauge interactions.}
\end{figure}

The total cross section reads:
\begin{equation}
\sigma=\left( \frac{g^2_{*}}{\Lambda^2}\right)^2 \frac{s}{4 \pi} \left( 1-\frac{m^{*2}}{s}\right)^2=\left( \frac{s}{\Lambda^2}\right)^2 \frac{4 \pi}{s} \left( 1-\frac{m^{*2}}{s}\right)^2 \, ,
\label{crossCont}
\end{equation}
where in the last step we set $g^{2}_{*}=4 \pi$, again keeping with the usual choice found in the literature \cite{Baur:1989kv}. We remark here that $g_*$ is an effective strong coupling constant, analogous to the $\rho$-meson effective coupling $g_\rho^2/(4\pi)\approx 2.1$, arising from the new "meta-color" force exchanged between preon sub-constitutents. Contact interactions are then normalised choosing $g^{2}_{*}=4 \pi$~\cite{Eichten:1983hw}. While other normalisations would of course be possible we stick to the current standard choice mainly in order to be able to directly compare our results with those of the experimental analyses~\cite{Aad:2013jja,Chatrchyan:2013ac,Aad:2012ab} which do follow such normalisation. In eq.~(\ref{crossCont}) $m^{*}$ stands for the excited lepton mass and the Mandelstam variable $s$ for the square of the collider energy. In the case of contact interactions there is no angular dependence in the cross section. Therefore, we just write down the result for the integrated total cross section. The model embedding doubly charged leptons in the contact interaction framework has been implemented in CalcHEP in \cite{Leonardi:2014aa}. 
In Fig.~\ref{fig:crosssections} we compare explicitly the exotic doubly charged lepton production cross sections via both gauge and contact interactions for the process of interest $e^{-}e^{-} \rightarrow E^{--}\nu_{e}$. We notice that the production via contact interactions dominates the production via gauge interactions from one to two orders of magnitude (for example $\Lambda=5$ TeV and $m^{*}=0.5$ TeV for the two different center of mass energies $\sqrt{s}=1,3$ TeV). 
\begin{table*}[t]
\begin{ruledtabular}
\begin{tabular}{ccccc}&$\sigma^*_x$ (nm)&$\sigma^*_y $ (nm)&$\sigma^*_z$ ($\mu$m)& $N$\\\hline
ILC & 481 & 2.8 & 250& 1.74$\times 10^{10}$\\
CLIC & 45&1&44&$3.72 \times 10^9$
\end{tabular}
\end{ruledtabular}
\caption{\label{tab:ISRpar} Beam parameters used to implement the initial state radiation (ISR) and the beamstrahlung effects in our estimates of the signal and background~\cite{Djouadi:2007ik,Linssen:2012hp,Accomando:2004sz}. }
\end{table*}

\section{Standard Model background and LSD invariant mass distribution}
\label{Sec4}
In this section we consider the signatures produced by the decay of the unstable doubly charged lepton and we focus on a pure final lepton state by considering the decay $E^{--} \to W^{-} e^{-} \to e^{-} \ell^{-} \nu_{\ell}$. In particular, the decay cascade determines the following final state event
$e^{-}e^{-} \to e^{-} e^{-} \nu_{e}\bar{\nu}_{e}$, where we  only consider the electron flavour for sake of simplicity. 

Such final particle set may be obtained either via contact or gauge interactions involved in the excited lepton production and decay process. 
Therefore, there are four different production-decay combinations. As shown in \cite{Leonardi:2014aa}, the contact interactions mechanism is always dominant for the production of $E^{--}$ whereas the decay exhibits a dependence on the relative values of the parameters $(m^{*},\Lambda)$. In order to obtain results as general as possible, in this  study we generate the signal  taking into account all of the 8 Feynman diagrams describing the signal process. All the combinations of contact and gauge driven production and decay processes are therefore considered together with their interferences.

The final state, $e^{-} e^{-} \nu_{e}\bar{\nu}_{e}$, translates experimentally into a like-sign dilepton (LSD) and missing transverse energy ($\slashed{E}_{T}$) signature, due to the undetectable neutrino and anti-neutrino. In order to shape the detection strategy of such a signal, an estimate of the backgrounds in the $e^{-}e^{-}$ beam setting is necessary. We consider only the SM induced background by using the CalcHEP and MadGraph \cite{Alwall:2014hca} generators. The contribution of additional beyond SM physics is not taken into account in the present work. In order to discuss the background classification we adopt the base kinematic cuts in Tab.~\ref{Tab1}. They are  typically adopted  to reproduce a general purpose detector geometrical acceptance and minimal requirements to detect charged leptons. The condition $p_T^{max}(e^{-}) > 50 $ GeV is used as a trigger for new physics searches. Let us start with the SM irreducible backgrounds. 

The SM process $e^{-} e^{-}\to e^{-} e^{-} \nu_{e}\bar{\nu}_{e}$ is described by a total of 28 Feynman diagrams (including those due to the exchange of identical particles). The ClacHEP generator allows to compute all of them including the various interferences. In particular such interferences are numerically quite important as they turn out to be mainly destructive. A simple estimate of the dominating diagrams would sensibly overestimate this background. This process exhibits exactly the same particle content as the signal process.  The cross sections obtained according to the base kinematic cuts in Tab.~\ref{Tab1} are: $188.2$ fb at $\sqrt{s}=1$ TeV and $209.3$ fb at $\sqrt{s}=3$ TeV. 

The SM process $e^{-} e^{-}\to e^{-} e^{-} \nu_{e} \nu_{e} \bar{\nu}_{e}\bar{\nu}_{e} $ is described by a total of 301 Feynman diagrams (including those due to the exchange of identical particles). The CalcHEP generator does not allow to handle six particles in the final state with satisfying accuracy. For this reason we study this process by means of MadGraph~\cite{Alwall:2007st}. Despite the particle content is different from that of the signal, there are two additional neutrinos, it represents also an irreducible background being the four neutrinos understood as $\slashed{E}_T$. However we expect this process to be suppressed by the SM couplings relative to the process $e^{-} e^{-}\to e^{-} e^{-} \nu_{e}\bar{\nu}_{e}$. We consider the interferences among all the diagrams and the cross sections obtained according to the base kinematic cuts in Tab.~\ref{Tab1} are: $0.306$ fb at $\sqrt{s}=1$ TeV and $1.356$ fb at $\sqrt{s}=3$ TeV. Therefore this process accounts for a fairly tiny fraction of the irreducible background, namely smaller than the 1\%, when compared with the process $e^{-} e^{-}\to e^{-} e^{-} \nu_{e}\bar{\nu}_{e}$.

Let us briefly comment on the reducible backgrounds. According to our signal signature we look for like-sign dilepton and missing energy. Possible sources of reducible backgrounds are: either jet or photons misidentified with electrons and non-prompt leptons coming from heavy flavour hadrons. Standard experimental techniques allow to have typical jet rejection factors into fake electrons of the order of $10^{-5}$ for the ATLAS detector \cite{Aad:2009wy}. The new generation detectors at ILC and CLIC are expected to perform at least at same level. For example, let us consider the process $e^{-} e^{-} \to jet \, jet \, e^{-} \nu_{e}$. The cross section is $\mathcal{O}(100)$ fb. However once it is combined with the suppression factor of $10^{-5}$, the corresponding cross section safely becomes negligible with respect to the SM irreducible background $e^{-} e^{-} \to e^{-} e^{-} \nu_{e} \bar{\nu}_{e}$. Moreover the second jet has to be not reconstructed or merged with the first jet, determining a further reduction of the rate from this type of background. In summary we focus on the SM irreducible background $e^{-} e^{-}\to e^{-} e^{-} \nu_{e}\bar{\nu}_{e}$ that provides the dominant contribution. 

According to previous studies carried out in \cite{Biondini:2012ny,Leonardi:2014aa}, the LSD invariant mass is an appropriate kinematic distribution to look at to discriminate between the background and the signal. However, we study also different kinematic variables in order to improve the signal strength over the background by establishing useful cuts at the level of the CalcHEP generator.  
\begin{table*}
\begin{tabular}{c|c|c|c}
\hline
\hline
\multicolumn{4}{c}{ILC - $\sqrt{s}= 1 $ TeV}\\
\hline
\hline
Model & $\sigma$ (fb) & $\sigma_{\text{ISR}}$ (fb) & $\sigma_{\text{ISR}}$ / $\sigma$\\
\hline 
SM ($e^{-} e^{-}\to e^{-} e^{-} \nu_{e}\bar{\nu}_{e}$)$\phantom{x}$ & 35.2 & 29.6  &  $\approx 0.84$ \\
\hline
$\Lambda=10$, $m^{*}=0.5$ (TeV)  $\phantom{x}$ & 25.67 & 19.93  &  $\approx 0.78$ \\
\hline
$\phantom{x}\Lambda=10$, $m^{*}=0.65$ (TeV)  $\phantom{x}$ & 17.36 & 12.32  &  $\approx 0.71$ \\
\hline
$\Lambda=10$, $m^{*}=0.8$  (TeV) $\phantom{x}$ & 8.16 & 4.86  &  $\approx 0.60$ \\
\hline 
\hline
\end{tabular}
\hspace{1.2 cm}
\begin{tabular}{c|c|c|c}
\hline
\hline
\multicolumn{4}{c}{CLIC - $\sqrt{s}= 3 $ TeV}\\
\hline
\hline
Model & $\sigma$ (fb) & $\sigma_{\text{ISR}}$ (fb) & $\sigma_{\text{ISR}}$ / $\sigma$\\
\hline 
SM ($e^{-} e^{-}\to e^{-} e^{-} \nu_{e}\bar{\nu}_{e}$) $\phantom{x}$ & 78.8 & 40.6  &  $\approx 0.51$ \\
\hline
$\Lambda=15$, $m^{*}=1.5$ (TeV)$\phantom{x}$ & 48.35 & 8.28  &  $\approx 0.17$ \\
\hline
$\Lambda=15$, $m^{*}=2.0$ (TeV)$\phantom{x}$ & 31.64 & 4.42 &  $\approx 0.14$ \\
\hline
$\Lambda=15$, $m^{*}=2.5$ (TeV)$\phantom{x}$ & 13.98 & 1.52  &  $\approx 0.11$ \\
\hline 
\hline
\end{tabular}
\caption{\label{Tab4} The cross sections for the SM irreducible background and signal are displayed. They correspond to the Improved Kinematic Cuts as given in \ref{Tab2}. We consider different values for $m^{*}$ at fixed $\Lambda$. The effect of the ISR and beamstrahlung is shown and it is quite more important for the signal. The larger $m^{*}$ the larger the suppression.}
\end{table*}

\begin{figure*}[t]
\includegraphics[width=12.0cm]{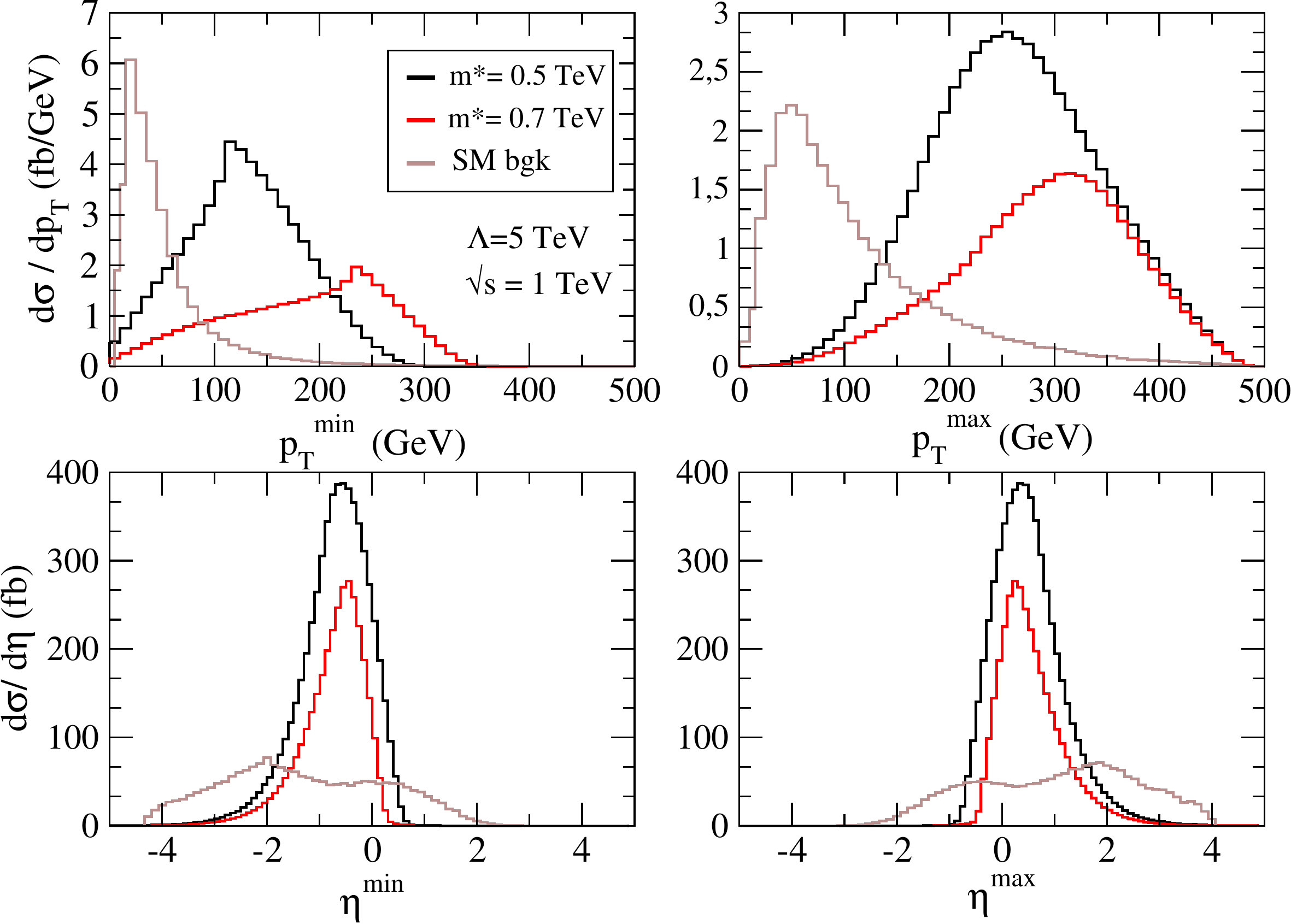}
\caption{\label{fig:distrBK_1} (Color online). The kinematic variables for the signal and the SM background  are shown. With $p_T^{max(min)}$ we label the highest (lowest) energetic electron transverse momentum distribution, and with $\eta^{max(min)}$ we label the highest (lowest) energetic electron pseudorapidity. The model parameters are $m^{*}=0.5,0.7 \, \text{TeV},\Lambda=5 \, \text{TeV}$. No cuts are implemented to obtain these distributions.}
\end{figure*}

\begin{figure}[h]
\includegraphics[width=7.25 cm]{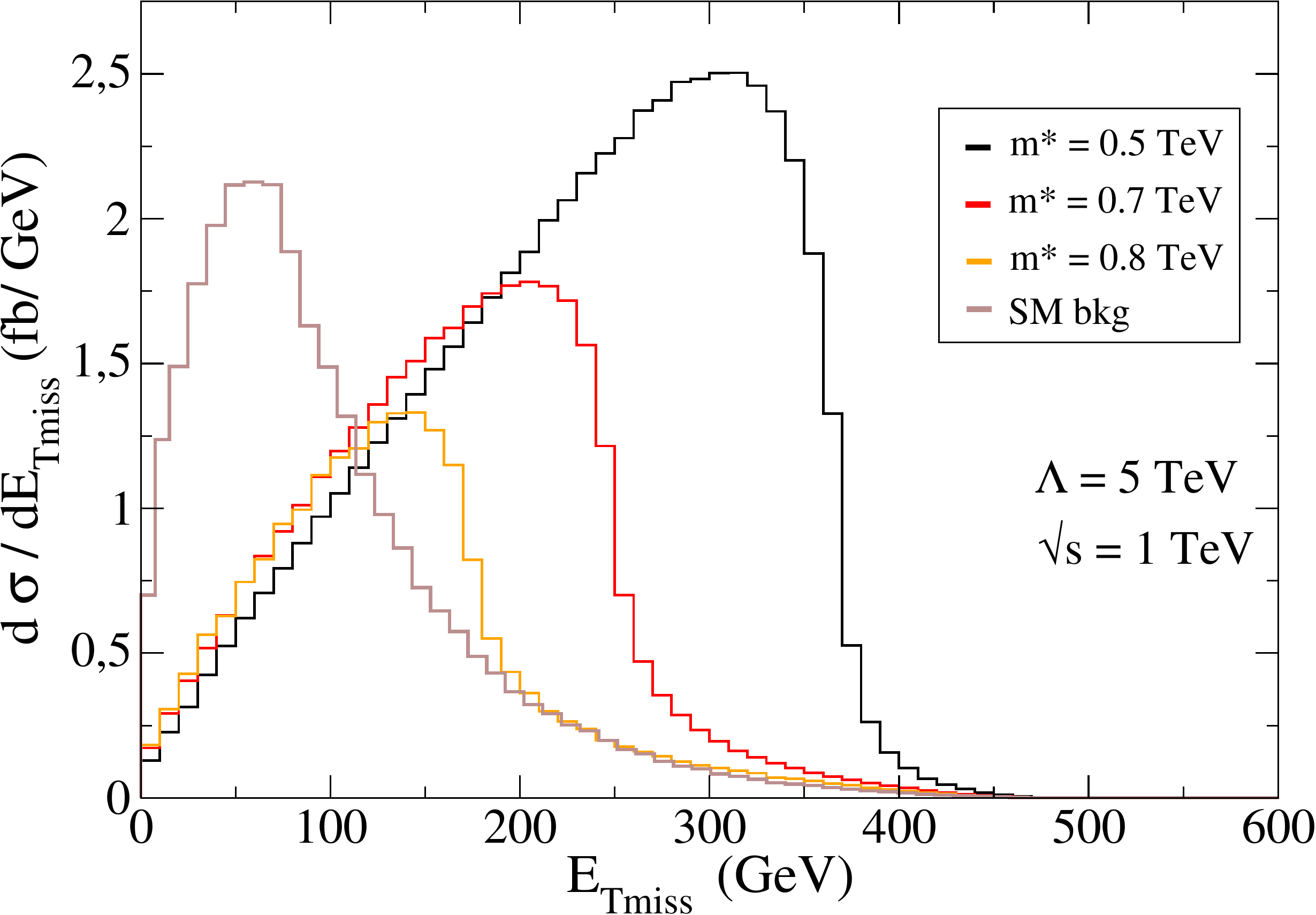}
\caption{\label{fig:missingT} (Color online). The missing transverse momentum of the undetectable neutrino and anti-neutrino is shown for the signal and the SM background processes. The model parameters are $m^{*}=0.5,\, 0.7,\, 0.8 \, \text{TeV},\Lambda=5 \, \text{TeV}$.}
\end{figure}
\begin{figure*}[t]
\includegraphics[width=12cm]{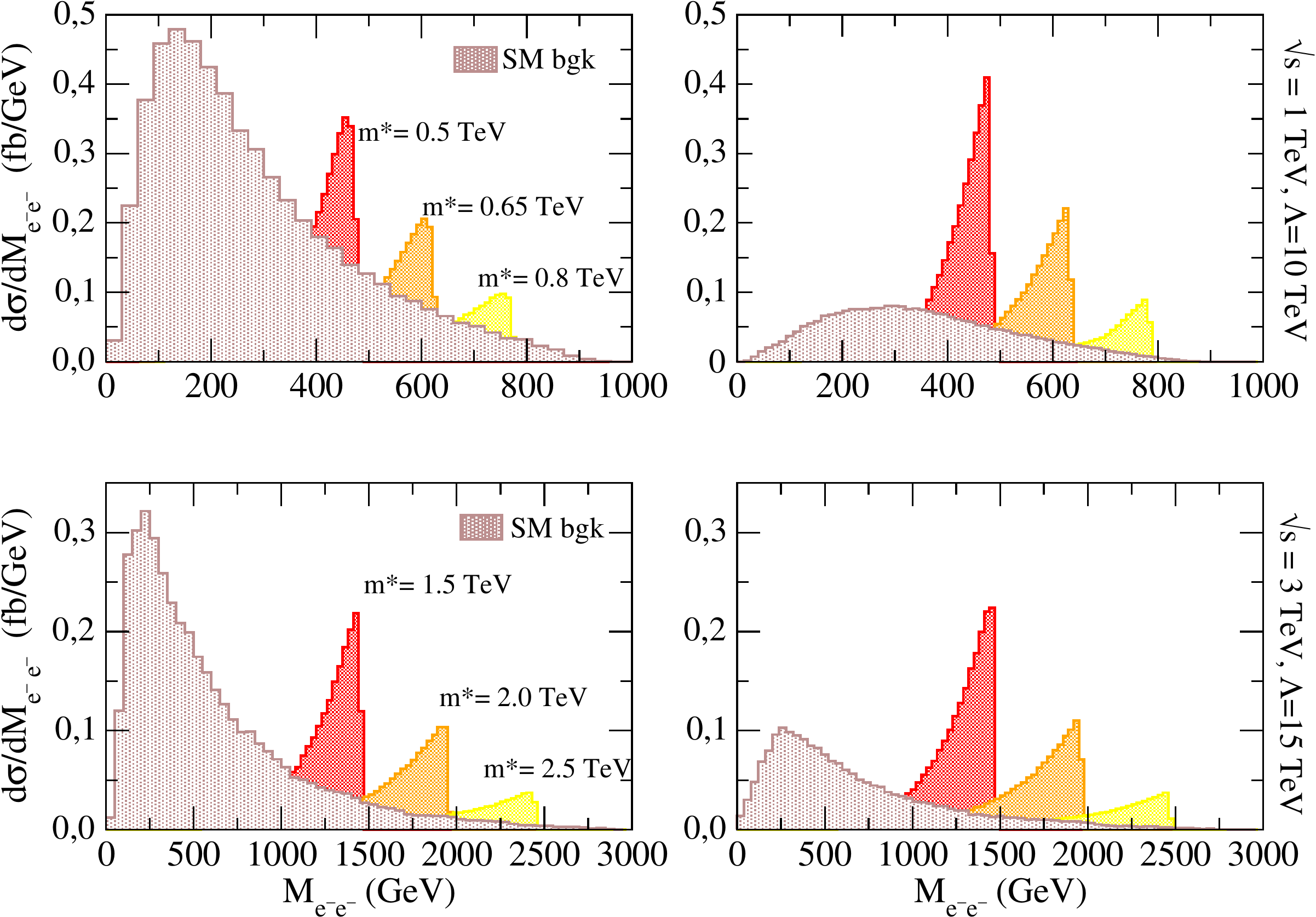}
\caption{\label{fig:invmass} (Color online). The LSD invariant mass distributions for both signal and background are displayed. In the top panels we show the case of ILC with $\sqrt{s}= 1$ TeV with the base kinematic cuts (left) and with the improved kinematic cuts (right). Signal distributions are shown for $\Lambda=10$ TeV  and different values of the doubly charged lepton mass $m^{*}=0.5, 0.65, 0.8 \; \text{TeV}$.  
The lower panels show the case of CLIC at $\sqrt{s}= 3$ TeV again with the base kinematic cuts (left) and with the improved kinematic cuts (right). Signal distributions are shown for $\Lambda=15$ TeV  and different values of the doubly charged lepton mass $m^{*}=1.5, 2.0, 2.5 \; \text{TeV}$.
}
\end{figure*}
\begin{figure*}[ht]
\includegraphics[width=12cm]{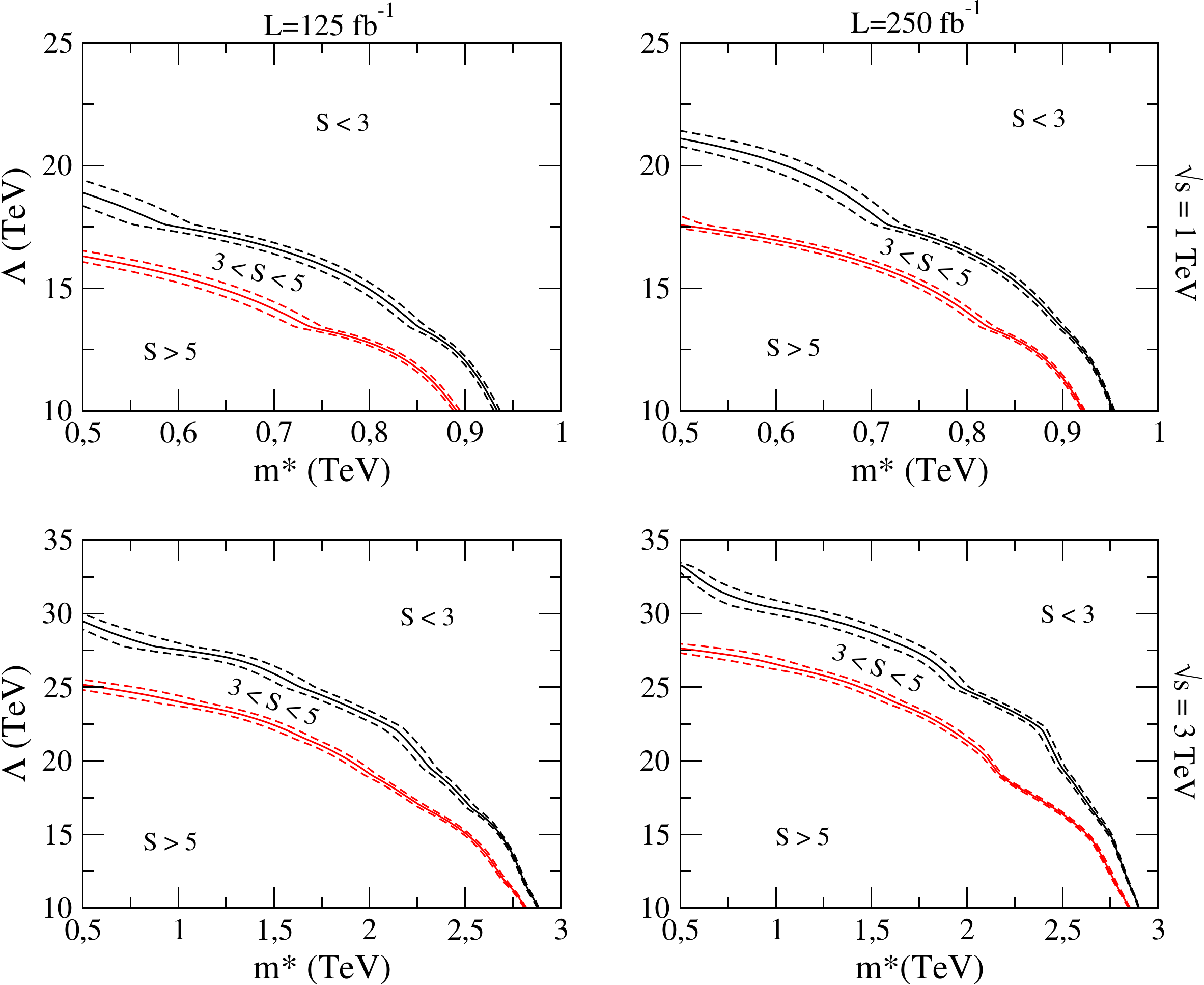}
\caption{\label{fig:lumicurves} 
(Color online) Contour plots of the statistical significance ($S$) at 3 and 5-sigma values, in the parameter space ($\Lambda, m^*$) within the statistical error on $S$ (dashed lines). The regions below the curves are excluded respectively at 3-sigma level (black line) or at the 5-sigma level (red line). We show both cases of $\sqrt{s}=1\, \text{TeV, upper panel and }\,\sqrt{s}=3$ TeV, lower panel for two different choices of the integrated luminosity $L=125$ fb$^{-1}$ and $L=250$ fb$^{-1}$. }
\end{figure*}

We study the differences in the kinematic distributions of signal and background by means of CalcHEP. 
We consider the following kinematic variables: most energetic electron transverse momentum and pseudo-rapidity, less energetic electron transverse momentum and pseudo-rapidity and missing transverse energy. We expect that the transverse energy distribution is peaked to higher values in the signal with respect to the background. Indeed, the neutrino produced in association with the excited lepton has to balance the heavy expected mass $m^{*}$ in the transverse plane. In Fig.~\ref{fig:distrBK_1}  we show the electrons transverse momentum and pseudo-rapidity distributions for the SM background against the same distribution for the signal for two particular choices of points in the parameter space ($m^{*}=0.5\, \text{and}\, 0.7$ TeV, $\Lambda=5$ TeV). We see that a cut on  $p_T^{\text{max}}$, the transverse momentum of the electron with leading $p_T$, of the order of $200$ GeV would sensibly suppress tha SM background relative to the signal. 
The asymmetry of the pseudo-rapidity distributions suggests to adopt  an asymmetric cut on $\eta^{\text{max}}$, the pseudo-rapidity of the highest energetic electron, 
and therefore we apply: $-1<\eta^{\text{max}} <2.5$ . 
The result for the missing transverse energy, $\slashed{E}_T$, is shown in Fig.~\ref{fig:missingT}, where the signal distribution exhibits the maximum at higher values with respect to the SM background. We find that the $\slashed{E}_T$ distribution  helps to device a useful cut to suppress the SM background  relative to the signal. We clearly see that a cut on the missing transverse energy, $\slashed{E}_T > 100$ GeV, should suppress the the SM background more than the signal.
The base kinematic cuts and the improved kinematics cuts to disentangle the signal from the background are summarised respectively in Tab. \ref{Tab1} and \ref{Tab2}.   

In Fig.~\ref{fig:invmass} we show the invariant LSD mass distribution for the signal and the background. Different values of the parameters $(m^{*},\Lambda)$ are considered. In the upper panels we consider the case of ILC at $\sqrt{s}=1$ TeV with $\Lambda =10 $ TeV and $m^*=0.5,0.65, 0.8$ TeV, with the base cuts of Tab.~\ref{Tab1} (left) and the improved kinematic cuts of Tab.~\ref{Tab2} (right). The same is shown for CLIC at $\sqrt{s}=3$ TeV with $\Lambda =15 $ TeV and $m^*=1.5,2.0, 2.5$ TeV. In all cases the correlation of the peak of the LSD invariant mass  distribution with the value of $m^*$ is quite evident.

Finally we study the statistical significance for the signal at 3 and 5-sigma level. 
We address this aspect in order to assess the accessible parameter space $(m^{*},\Lambda)$ at linear colliders, at least in this particular beam setting ($e^-e^-$) and at the level of ideally reconstructed particles. However, we point out that our final numerical simulations include  both for the signal and SM background the effects of initial state radiation and beamstrahlung~\cite{Djouadi:2007ik,Linssen:2012hp}.  These effects are taken into account by the CalcHEP generator by specifying specific beam characteristics such as the dimension of the bunch along the beam direction ($\sigma^*_z$), the dimension of the beam in the transverse plane ($\sigma^*_x$ and $\sigma^*_y$) and the number of particles per bunch ($N$). The specific values used for estimating the ISR and beamstrahlung at ILC and CLIC are given in Tab.~\ref{tab:ISRpar}. 
These effects turn out to be more important for the signal than for the SM background. This is understood by considering that radiation effects will tend to degrade the high energy peak of the beam spectrum. Thus we expect these effects to reduce more the signal cross section, say for $m^*\approx 1 $ TeV, than the SM background cross-section where one has to produce particles with mass of the order of the electroweak scale ($M_W, M_Z$).  Indeed for the signal the combined effect of ISR and beamstrahlung is more important as the mass $m^*$ of the exotic leptons is increased (see Tab.~\ref{Tab4}).

At a given integrated luminosity $L$ and for any point in the parameter space $(m^{*},\Lambda)$, we can provide the expected number of events for the signal and the background as follows:
\begin{equation}
N_s=L \sigma_s \, , \quad N_b=L  \sigma_b \, ,
\label{events}
\end{equation}
where the the cross section for the signal, $\sigma_s$, and the one for the background, $\sigma_b$, are provided by the CalcHEP generator according to the improved kinematic cuts given in Tab.~\ref{Tab2}. We use the following definition for the statistical significance, that reads
\begin{equation}
S=\frac{N_s}{\sqrt{N_b+N_s}} \, .
\label{SS}
\end{equation} 
The statistical significance is then a function of two parameters, namely $m^{*}$ and $\Lambda$ through the expected number of events for the signal and the background in (\ref{events}). 
Performing the scanning over the parameter space we can derive the experimental evidence region ($S \geq 3$) and experimental discovery region ($S \geq 5$). 
The results are shown in Fig.~\ref{fig:lumicurves}. We have also included an estimate of the statistical error on the significance $S$ by propagating in quadrature the statistical errors in $N_s$ ($\sqrt{N_s}$) and $N_b$ ($\sqrt{N_b}$). This gives rise to the  bands in the 3 and 5-sigma contour plots of the significance shown in Fig.~\ref{fig:lumicurves} by the dashed lines. We remark here that these results are based on a numerical computation which includes all 28 Feynman diagrams contributing to the dominant standard model process generated by CalcHEP, including all interferences. It turns out that the interferences are largely destructive and thus they make the SM background lower than it could be estimated by simply considering the dominant diagrams. We consider two different cases for the luminosity, $L=125$ fb$^{-1}$ and $L=250$ fb$^{-1}$, and two energies of the colliding electrons, $\sqrt{s}=1$ TeV and $\sqrt{s}=3$ TeV. This choice possibly corresponds to the nominal luminosity $L=500$ fb$^{-1}$ ($L=1000$ fb$^{-1}$) and energy for the ILC \cite{Djouadi:2007ik} (and CLIC \cite{Linssen:2012hp,Accomando:2004sz}) facilities. Indeed it has been argued~\cite{Schulte:2003aa} that the luminosity of the $e^-e^-$ collider should be $1/4$ of  the $e^+e^-$ configuration.
Accordingly  Fig.~\ref{fig:lumicurves} shows the phase space regions accessible by the ILC and CLIC at 3- and 5-$\sigma$ level with both values of the luminosity given above (125 fb$^{-1}$ and 250 fb$^{-1}$). Finally in Fig.~\ref{fig:comparison} the ILC and CLIC exclusion regions at 3-$\sigma$ are compared with bounds provided by the LHC analyses of run I at $\sqrt{s}=7,8$ TeV~\cite{Aad:2013jja,Chatrchyan:2013ac,Aad:2012ab,CMS:2015jga} and with the prospected exclusion bound on the parameter space ($\Lambda,m^*$) from a theoretical study of the three-lepton signature of the doubly charged excited electron at run II LHC ($\sqrt{s}=14$ TeV)~\cite{Leonardi:2014aa}.  For example, for a mass of $m^{*} = 0.6$ TeV the corresponding limit on the compositeness scale at ILC (CLIC) would be $\Lambda>18$ TeV ($\Lambda>28$ TeV) at 3-sigma level and for $L=125$ fb$^{-1}$, instead of $\Lambda > 15$ TeV at the LHC (run I)~\cite{CMS:2015jga}.
The situation appears to be even more encouraging at larger values of the excited lepton mass. For $m^*\approx 2$ TeV the expected CLIC exclusion bound on $\Lambda$ would be (see Fig.~\ref{fig:lumicurves} and Fig.~\ref{fig:comparison}) $\Lambda > 22 $ TeV ($\Lambda > 25 $ TeV)  for L=125 fb$^{-1}$ (L=250 fb$^{-1}$), which is much better than the current (run I) LHC bound: $\Lambda > 5 $ TeV~\cite{CMS:2015jga} (Fig.~\ref{fig:comparison}). 
We note that the CLIC bound on $\Lambda$  for  $m^*\approx 2$ TeV ($\Lambda >22$-$25$ TeV)  fares pretty well also with the predictions about the same model at run II ($\sqrt{s}=14$ TeV) of LHC from ref.~\cite{Leonardi:2014aa} where for $m^*\approx 2$ TeV one expects, with an integrated luminosity of $300$ fb$^{-1}$, the lower bound on the compositeness scale $\Lambda > 11.6$ TeV. We remark also that the actual bounds quoted from the current run I LHC analyses~\cite{Aad:2013jja,Chatrchyan:2013ac,Aad:2012ab,CMS:2015jga} refer to excited leptons of ordinary iso-spin doublets and strictly speaking do not apply to the exotic charge states of the extended iso-spin multiplets, object of this work. Given the similarities of the interactions we do not expect the bounds from eventual dedicated search analyses of the exotic charge states to differ very much form those being currently considered.\\    
We conclude this section with an important remark. While we have not performed a simulation of the detector reconstruction efficiencies we expect our predictions for the contour plots of the statistical significance to be rather conservative.   Indeed from Fig.~\ref{fig:distrBK_1} we see that the signal electrons are more energetic than those of the background. The pseudo-rapidity distribution  is quite central for the signal (practically contained in the geometrical acceptance of the detector $-2.5<\eta<2.5$) whereas it is more spread out for the background. While the background electrons out of the geometrical acceptance are lost those that are within the geometrical acceptance are in any case less energetic than the ones from the signal. These elements favour a higher reconstruction efficiency for the electrons of the signal  than for those of the SM background leading thus to a higher statistical significance. We expect therefore that including such detector reconstruction efficiencies would increase the statistical significance and thus would enlarge the region of the parameter space that could be probed by the Linear Collider.
Overall we can say that while the ILC is obviously limited in the mass range by kinematics and could only fare better for lower masses only with higher luminosities than those considered here, the CLIC collider at $3$ TeV is highly competitive with LHC run II well within the region of high masses (see Fig.~\ref{fig:comparison} right panel).

\section{Conclusions and Outlook}
\label{Sec5}
In the paper we discuss the production of heavy exotic doubly charged leptons in the framework of forthcoming linear colliders. In particular we focus on the beam setting $e^{-} e^{-}$, that allows for the production of a single doubly charged lepton instead of the corresponding pair production. Since the excited doubly charged lepton mass is expected to be heavy, the $e^{-}e^{-}$ beam configuration determines a wider mass parameter space to explore with respect to the $e^{+}e^{-}$ beam setting. 

We consider the production of the heavy exotic lepton both within contact and gauge interactions. In the latter case we study in some detail the production cross section in relation with the different isospin multiplets ($I_W=1$ and $I_W=3/2$). In particular, we take into account the interference between the $t$ and $u$ channels and we provide the expression for the differential cross sections. The interference effect has not been previously evaluated and it is there for $E^{--} \in I_W=3/2$. Despite the production via contact interaction mechanism dominates over the gauge interaction one, the additional $t$-$u$ interference term produces a sizable contribution to the cross section in the gauge mediated case. It amounts to a reduction of the production cross section by a factor of one third at $\sqrt{s}=1$ TeV.

By considering the decay cascade $E^{--} \to W^{-}e^{-} \to \bar{\nu}_e e^{-}e^{-}$, the following final state particle set is studied: $ e^{-}e^{-}\bar{\nu}_e \nu_e$. Experimentally it translates into a LSD pair and missing transverse energy.  We study the LSD invariant mass distribution to assess the detection strategy of the signal in the linear collider and $e^{-}e^{-}$ beam setting. While we do not take into account other new physics sources besides the excited doubly charged lepton,  we study in detail the relevant SM background (28 Feynman diagrams and relative interferences). We perform an analysis of their kinematic distributions by means of CalcHEP including in the final simulations ISR and beamstrahlung effects both for the signal and the SM backgorund.  We then suggest a sample of kinematic cuts to discard background events whereas keeping the signal strength still high. We derive by means of CalcHEP the number of expected signal and background events taking into account some improved cuts. Therefore, we provide the 3 and 5-sigma statistical significance exclusion curves in the $(m^{*},\Lambda)$ parameter space. 

CalcHEP allows to study kinematic distributions for particles which are ideally reconstructed. In order to better establish the possibility to observe such exotic leptons at liner colliders, it would be desirable to introduce the effects of detectors like the efficiency reconstruction of the physical objects. However this effect, if taken properly into account, will work in favor of the signal over background event ratio, and so the statistical significance is expected to increase, in view of the fact that hard and central electrons, which are expected to be more frequent in the signal than in the SM background, are more easily reconstructed in a general purpose detector. Therefore  we expect that a detector simulation of the reconstruction efficiencies at the linear collider is bound to improve the region of the parameter space that can be probed at 3 and 5-sigma level relative to the current results shown in Fig.~\ref{fig:lumicurves}. We emphasize however that in the present study we have taken in to account basic detector effects like the geometrical acceptance (see Table~\ref{Tab1}).

Finally, we can say that based on the result presented here further investigations of the present model at  linear colliders are strongly encouraged perhaps investigating possible effects due to the available polarisation of the beams that  could still improve the bounds of Fig.~\ref{fig:lumicurves}. Also, at a linear collider it may still prove  interesting to investigate interference effects of the exotic doubly charged lepton exchanged as a virtual particle. For instance in $e^+e^- \to W^+W^-$, as opposed to the direct production. 

\begin{figure*}[t]
\includegraphics[width=14cm]{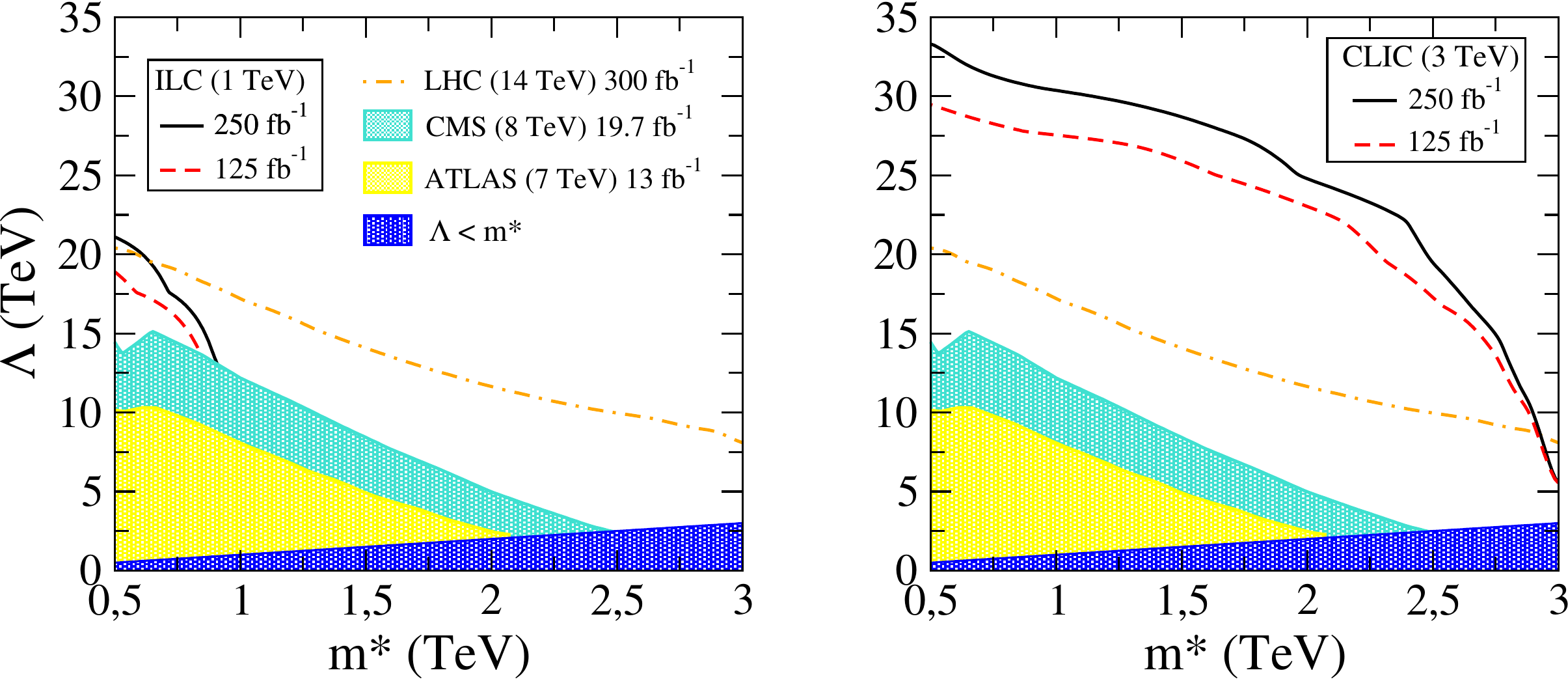}
\caption{\label{fig:comparison} 
(Color online) Comparison of the exclusion curves at 3-sigma level in the parameter space ($\Lambda, m^*$) from ILC (left) and CLIC (right), for L=250 fb$^{-1}$ solid (black) line, L=125 fb$^{-1}$ dashed (red) line, with the bounds from run I of LHC, (ATLAS 7 TeV 13 fb$^{-1}$~\cite{Aad:2013jja}, clear (yellow) shaded area, CMS 8 TeV \@19.7 fb$^{-1}$~\cite{CMS:2015jga}, light (blue) shaded area) and with the prospected bounds from run II of the LHC ($\sqrt{s}=14$ TeV, 300 fb$^{-1}$)~\cite{Leonardi:2014aa}, dot-dashed (orange) line. The dark (blue) shaded area is the region of the parameter space not allowed by requiring the internal consistency of the model, or that $m^*\le \Lambda$. Thus the points in the dark (blue) shaded region $\Lambda < m^*$ describe models which are not viable.
}
\end{figure*}

If there will be no direct observation of contact interactions and/or excited fermions at the LHC run II in the region of large  excited lepton masses, $m^* \approx 2 $ TeV, the CLIC lower bounds $\Lambda>22$-$25$ TeV, on the basis of  expectations of signatures of the same model at the LHC $\sqrt{s}=14$ TeV~\cite{Leonardi:2014aa}, remains highly  competitive with the bound expected for  run II of the LHC: $\Lambda > 11.6$ TeV  (see Fig.~\ref{fig:comparison} orange line).  
 
\begin{acknowledgments}
The authors acknowledge useful discussions with L. Fan\`{o}, R. Leonardi, F. Romeo and M. Szalay. 
This work was  in part supported by the research grant QUASAP of the Istituto Nazionale di Fisica Nucleare (INFN). 
\end{acknowledgments} 


%

\end{document}